# Fractional Differential Forms

by

Kathleen Cottrill-Shepherd [a)] and Mark Naber [b)]

Department of Mathematics

Monroe County Community College

Monroe, Michigan

48161-9746

**ABSTRACT**

A generalization of exterior calculus is considered by allowing the partial derivatives in the exterior derivative to assume fractional orders. That is, a fractional exterior derivative is defined. This is found to generate new vector spaces of finite and infinite dimension, fractional differential form spaces. The definitions of closed and exact forms are extended to the new fractional form spaces with closure and integrability conditions worked out for a special case. Coordinate transformation rules are also computed. The transformation rules are different from those of the standard exterior calculus due to the properties of the fractional derivative. The metric for the fractional form spaces is given, based on the coordinate transformation rules. All results are found to reduce to those of standard exterior calculus when the order of the coordinate differentials is set to one.

[a)] KShepherd@mail.monroe.cc.mi.us

[b)] MNaber@mail.monroe.cc.mi.us

## I. INTRODUCTION

In recent years exterior calculus has been generalized by basing it on various graded algebras, see for example Refs. 1 and 2. Other attempts at generalization are based on non-associative geometries, see for example Refs. 3 and 4. In this paper another attempt at generalization is made using fractional derivatives in the definition of the exterior derivative. That is, a fractional exterior derivative is defined. Having a fractional exterior derivative gives rise to the notion of coordinate differentials of fractional order. These in turn can be used to define vector spaces of fractional differential forms. This formalism is found to produce an infinite number of finite and infinite dimensional vector spaces associated with each point $P \in E^n$ (n dimensional Euclidean space).

In sections 2 and 3 a brief review of exterior calculus and fractional calculus will be given to fix notation and provide a convenient reference. They are by no means complete, but are sufficient for the purposes of this paper. Section 4 defines fractional form spaces based on the fractional exterior derivative. Basis sets are given for the new vector spaces and notation is fixed. In section 5 the definitions of closed and exact are expanded to include the fractional form case. Once defined, the notions of closed and exact forms are examined for these new vector spaces. In both cases the results reduce to those found in standard exterior calculus when the order of the coordinate differentials is set equal to one. Coordinate transformations are worked out for the fractional form spaces in section 6. The transformation rules are somewhat more complicated than for standard exterior calculus. They do, however, reduce to the usual transformation rules when the order or the coordinate differentials is set to one. Having found the coordinate transformation rule, a metric for the fractional form spaces is constructed. Metrical properties of these new vector spaces will be investigated in a later paper.

The convention in the literature is that the coordinate index is a super-script. For the topics presented in this paper it is more convenient for the coordinate index to be a sub-script rather than the traditional super-script. To avoid confusion with this, the summation convention will not be used in this paper.

## II. BRIEF REVIEW OF DIFFERENTIAL FORMS

The calculus of differential forms is an elegant branch of pure mathematics and a powerful tool in applied mathematics. A clear introduction to the field, with emphasis on applications, is given in Flanders [5]. Vector spaces at a point $P \in E^n$ (n dimensional Euclidean space) can be constructed out of expressions of the following type

$$\text{one forms, } \alpha = \sum_{i=1}^{n} a_i dx_i, \qquad (1)$$

$$\text{two forms, } \beta = \sum_{i,j=1}^{n} b_{ij} dx_i \wedge dx_j, \qquad (2)$$

$$\text{M}$$

$$\text{'n' forms, } \omega = w\, dx_1 \wedge dx_2 \wedge \text{L} \wedge dx_n. \qquad (3)$$

Where the $\{x_i\}$ are the Cartesian coordinates of $E^n$. The above sums are taken over all possible values of the indices with the constraint that

$$dx_i \wedge dx_j = -dx_j \wedge dx_i. \qquad (4)$$

The functions $a_i$, $b_{ij}$, *etc.* depend only on P and may be real or complex depending on the application. If a k-form, $\gamma$, is multiplying an m-form, $\mu$, the following would be true,

$$\gamma \wedge \mu = (-1)^{km} \mu \wedge \gamma. \tag{5}$$

The result would be zero if k + m > n. The exterior product, $\wedge$, is distributive, associative, and anti-symmetric. The dimension of the vector space of k-forms over $P \in E^n$ is $\binom{n}{k} = \frac{n!}{k!(n-k)!}$, which is zero if k > n. For the purposes of this paper let $F(k,k,n)$ denote the vector space of k-forms over $P \in E^n$. The apparently redundant 'k' in the above notation will be needed later for the fractional form case, as there is some additional freedom.

The exterior derivative is defined as,

$$d = \sum_{i=1}^{n} dx_i \frac{\partial}{\partial x_i}. \tag{6}$$

The exterior derivative maps k-forms into k+1-forms and has the following algebraic properties. Let $\gamma$ and $\lambda$ be k-forms, and $\mu$ be an m-form, then

$$d(\gamma + \lambda) = d\gamma + d\lambda, \tag{7}$$

$$d(\gamma \wedge \mu) = (d\gamma) \wedge \mu + (-1)^k \gamma \wedge d\mu, \tag{8}$$

$$d(d\gamma) = 0. \tag{9}$$

The last identity is called the Poincaré lemma. A form, $\gamma$, is called closed if $d\gamma = 0$. A form, $\gamma$, is called exact if there exists a form, $\mu$, such that $d\mu = \gamma$. The order of $\mu$ is one less than the order of $\gamma$. Exact forms are always closed. Closed forms are not always exact. The interested reader should consult Flanders [5] or Lovelock and Rund [6] for further details.

## III. BRIEF OVERVIEW OF FRACTIONAL CALCULUS

There are many books that develop fractional calculus and the various definitions of fractional integration and differentiation. The reader should consult Refs. 7, 8, and 9 for further details and applications. For the purposes of this paper the Riemann-Liouville definition of fractional integration and differentiation will be used. $\Gamma(q)$ is the gamma function (generalized factorial) of the parameter 'q' (i.e. $\Gamma(n+1) = n!$ for all whole numbers, 'n').

$$\frac{\partial^q f(x)}{(\partial(x-a))^q} = \frac{1}{\Gamma(-q)} \int_a^x \frac{f(\xi)d\xi}{(x-\xi)^{q+1}}, \quad \text{Re}(q) < 0, \tag{10}$$

$$\frac{\partial^q f(x)}{(\partial(x-a))^q} = \frac{\partial^n}{\partial x^n} \left[ \frac{1}{\Gamma(n-q)} \int_a^x \frac{f(\xi)d\xi}{(x-\xi)^{q-n+1}} \right] \quad \frac{\text{Re}(q) \geq 0}{n > q \text{ (n is whole)}} \tag{11}$$

The parameter 'q' is the order of the integral or derivative and is allowed to be complex. Positive real values of 'q' represent derivatives and negative real values represent integrals. Equation (10) is a fractional integral and equation (11) is a fractional derivative. In this paper only real and positive values of 'q' will be considered. Notice that the derivative written in this form becomes a non-local object. Fractional derivatives have many interesting properties. For example the derivative of a constant need not be zero (the initial point 'a' in the above definition is set to zero in the following).

$$\frac{\partial^q 1}{(\partial x)^q} = \frac{x^{-q}}{\Gamma(1-q)} \tag{12}$$

The derivative of powers of x is

$$\frac{\partial^q x^p}{(\partial x)^q} = \frac{\Gamma(p+1)}{\Gamma(p-q+1)} x^{p-q}, \quad \begin{array}{c} p > -1 \\ q \geq 0 \end{array}. \tag{13}$$

Composition and product rules for fractional derivatives are given below. In the following, 'n' is a whole number and 'q' is a complex number whose real part is greater than zero.

$$\frac{\partial^n}{\partial x^n} \frac{\partial^q}{(\partial(x-a))^q} f(x) = \frac{\partial^{n+q}}{(\partial(x-a))^{n+q}} f(x) \tag{14}$$

$$\frac{\partial^q}{(\partial(x-a))^q} \frac{\partial^{-q}}{(\partial(x-a))^{-q}} f(x) = f(x) \tag{15}$$

$$\frac{\partial^{-q}}{(\partial(x-a))^{-q}} \frac{\partial^q}{(\partial(x-a))^q} f(x) \neq f(x) \tag{16}$$

Composing derivatives where both have fractional order is given by the following formula.

$$\frac{\partial^p}{(\partial(x-a))^p} \frac{\partial^q}{(\partial(x-a))^q} f(x) = \frac{\partial^{p+q}}{(\partial(x-a))^{p+q}} f(x) - \sum_{j=1}^{k} \frac{\partial^{q-j}}{(\partial(x-a))^{q-j}} f(x) \bigg|_{x=a} \frac{(x-a)^{-p-j}}{\Gamma(1-p-j)}$$

(17)

Where, $0 \leq k-1 \leq q \leq k$, $p \geq 0$, and k is a whole number. The product rule is,

$$\frac{\partial^q}{(\partial x)^q}(fg) = \sum_{j=0}^{\infty} \binom{q}{j} \left(\frac{\partial^{q-j} f}{(\partial x)^{q-j}}\right) \left(\frac{\partial^j g}{\partial x^j}\right)$$

(18)

The above formula and definitions can be found in Refs. 7, 8, and 9.

## IV. FRACTIONAL FORM SPACES

If the partial derivatives in the definition of the exterior derivative are allowed to assume fractional orders, a fractional exterior derivative can be defined.

$$d^v = \sum_{i=1}^{n} dx_i^v \frac{\partial^v}{(\partial(x_i - a_i))^v}.$$

(19)

Note that the subscript 'i' denotes the coordinate number, the superscript '$v$' denotes the order of the fractional coordinate differential, and $a_i$ is the initial point of the derivative.

Sometimes the notation $\partial_i^v$ will be used to denote $\dfrac{\partial^v}{(\partial(x_i - a_i))^v}$.

In two dimensions (x,y), the fractional exterior derivative of order $v$ of $x^p$, with the initial point taken to be the origin, is given by

$$d^v x^p = dx^v \frac{\Gamma(p+1)}{\Gamma(p-v+1)} x^{p-v} + dy^v \frac{x^p}{y^v \Gamma(1-v)}. \tag{20}$$

For specific values of the derivative parameter the following results are obtained.

$$v = 0 \quad d^0 x^p = 2x^p \tag{21}$$

$$v = 1 \quad d^1 x^p = dx^1 p x^{p-1} \tag{22}$$

$$v = 2 \quad d^2 x^p = dx^2 p(p-1) x^{p-2} \tag{23}$$

By analogy with standard exterior calculus, vector spaces can be constructed using the $dx_i^v$. Let F($v$,m,n) be a vector space at $P \in E^n$. '$v$' denotes the sum of the fractional differential orders of the basis elements, 'm' denotes the number of coordinate differentials appearing in the basis elements, 'n' the number of coordinates, and $\{x_i\}$ are the Cartesian coordinates for $E^n$. For example, a basis set for $F(v,1,n)$ would be $\{dx_1^v, dx_2^v, \ldots, dx_n^v\}$ and an arbitrary element of $F(v,1,n)$ would be expressed as

$$\alpha = \sum_{i=1}^{n} \alpha_i \, dx_i^v. \tag{24}$$

For a fixed $v$ this is an 'n' dimensional vector space. Also note that there is a different vector space for each value of $v$. For $v = 1$ the one forms from exterior calculus are recovered. Now suppose that the basis elements are made up of two coordinate differentials, $F(v,2,n)$. In this case the basis set is more complicated.

$$\left\{ dx_1^{\mu_{11}} \wedge dx_1^{\mu_{21}},\ dx_1^{\mu_{11}} \wedge dx_2^{\mu_{31}},\ \ldots,\ dx_n^{\mu_{n-1m}} \wedge dx_n^{\mu_{nm}} \,\middle|\, \mu_{ij} + \mu_{kj} = v \right\} \tag{25}$$

Note that $dx_1^{\mu_{11}} \wedge dx_1^{\mu_{21}}$ would be zero if and only if $\mu_{11} = \mu_{21}$, etc. An arbitrary element of $F(v,2,n)$ would be expressed as a sum of the form

$$\beta = \sum_{i=1}^{n} \sum_{j=1}^{n} \int_0^v \left( \beta_{ij}(v_i, v - v_i) dx_i^{v_i} \wedge dx_j^{v-v_i} \right) dv_i, \tag{26}$$

where $\beta_{ii}(\mu,\mu) = 0$. Unlike the previous vector space, $F(v,1,n)$, this is clearly infinite dimensional for any value of $v$. Not only is it infinite but it is uncountably infinite. An arbitrary element of $F(v,3,n)$ would be expressed as an integral of the form

$$\beta = \sum_{i=1}^{n}\sum_{j=1}^{n}\sum_{k=1}^{n} \int_0^v \int_0^{v-v_i} \left( \beta_{ijk}(v_i, v-v_j, v-v_i-v_j) dx_i^{v_i} \wedge dx_j^{v_j} \wedge dx_k^{v-v_i-v_j} \right) dv_j dv_i. \tag{27}$$

With each step up on the middle index of $F(v,m,n)$ another integral and summation is included. A basis set for this vector space would be

$$F(v,m,n) = \left\{ dx_{i_1}^{\mu_{i_1 1}} \wedge dx_{i_2}^{\mu_{i_2 1}} \wedge \ldots \wedge dx_{i_m}^{\mu_{i_m 1}},\ \ldots \,\middle|\, \sum_{k=1}^{m} \mu_{i_k j} = v \right\}. \tag{28}$$

The basis elements range over all possible combinations of the fractional coordinate differentials and all possible choices for the $\mu's$. Note that 'm' need not be less than or equal to 'n'.

Let $P \in E^n$, and let $A \in F(v, m, n)$ and $B \in F(\mu, k, n)$ at the point P. Then the exterior product of A and B maintains the antisymmetry property of equation (5),

$$A \wedge B = (-1)^{km} B \wedge A \in F(\mu + v, k + m, n). \tag{29}$$

If $k + m > n$, $A \wedge B$ need not be zero. Equation (7) is also maintained due to the linearity of the fractional derivative. Equation (8) is not maintained due to the product rule for the fractional derivative (see equation (18)). Note also that $d^v$ maps $F(\mu, k, n)$ into $F(\mu + v, k + 1, n)$.

## V. CLOSED AND EXACT FRACTIONAL FORMS

By analogy with exterior calculus the notions of closed and exact can be extended to fractional forms.

Let $g \in F(\mu, k, n)$ then g is $v$-exact if $\exists$ an $f \in F(\mu - v, k - 1, n)$ such that $d^v f = g$.

Let $g \in F(v, k, n)$ then g is $\mu$-closed if $d^\mu g = 0$.

To examine the notion of exactness (integrability conditions) in the fractional form case the kernel is needed for the fractional derivative operator. This will be denoted by, $Ker(\partial_i^v)$. In the following the initial point of the derivative will be taken to be the origin. Solve

$$\partial_i^\nu(h) = 0. \tag{30}$$

Equation (30) is solved using equation (11). Let 'm' be the first whole number greater than or equal to $'\nu'$ then

$$h = (x_i)^{\nu-m} \sum_{k=0}^{m-1} c_k (x_i)^k. \tag{31}$$

This is basically the result from Oldham[7] page 155. The $c_k's$ can be functions of the other coordinates. The kernel for the operator, $d^\nu$ (when restricted to act only on scalar functions) is similarly constructed, $Ker(d^\nu)$.

$$d^\nu f = \sum_{i=1}^m dx_i^\nu \frac{\partial^\nu f}{(\partial x_i)^\nu} = 0 \tag{32}$$

$$\Rightarrow f = \left(\prod_{i=1}^n x_i\right)^{\nu-m} \left(\sum_{k_1=0}^{m-1} \text{L} \sum_{k_n=0}^{m-1} C_{k_1 \text{L} k_n} (x_1)^{k_1} \text{L} (x_n)^{k_n}\right) \tag{33}$$

The $C_{k_1 \text{L} k_n}$ are now constants and 'm' is once again the first whole number greater than or equal to $'\nu'$.

The fractional integrability conditions can now be constructed for the following restricted case. Let g be a $\nu$ - form in $F(\nu,1,n)$

$$g = \sum_{i=1}^n \alpha_i \, dx_i^\nu. \tag{34}$$

When can a 0-form, f, be found such that $d^v f = g$? If such an f exists it will be contained in the family of functions given by

$$f = \partial_i^{-v}(\alpha_i) + (x_i)^{v-m} \sum_{k=0}^{m-1} c_k (x_i)^k. \tag{35}$$

Recall that there is no sum over the repeated indices, except where a summation symbol is encountered. This solution must satisfy, $\partial_j^v f = \alpha_j$,

$$\partial_j^v \left( \partial_i^{-v}(\alpha_i) + (x_i)^{v-m} \sum_{k=0}^{m-1} c_k (x_i)^k \right) = \alpha_j. \tag{36}$$

The above equation is to be solved to determine the unknown functions $c_k$, and must be true for all values of 'i'. Equation (36) can be rearranged to give

$$\sum_{k=0}^{m-1} \left( \partial_j^v c_k \right)(x_i)^k = \frac{\alpha_j - \partial_j^v \left( \partial_i^{-v} \alpha_i \right)}{(x_i)^{v-m}}. \tag{37}$$

The left hand side of (37) is a polynomial of order 'm-1' in the variable $x_i$, hence it can only be solved for the $c_k$ if the following is true

$$\frac{\partial^m}{\partial x_i^m} \left( \frac{\alpha_j - \partial_j^v \left( \partial_i^{-v} \alpha_i \right)}{(x_i)^{v-m}} \right) = 0. \tag{38}$$

The above equation is the integrability condition for fractional forms of type $F(v,1,n)$.

Note that if $v = m = 1$ the usual integrability conditions from exterior calculus are recovered.

Consider the second definition and examine what it takes to be closed for fractional forms.

Let $\alpha = \sum_{i=1}^{n} \alpha_i \, dx_i^v \in F(v,1,n)$ and consider its fractional exterior derivative.

$$d^\mu \alpha = \sum_{i=1}^{n} d^\mu \left( \alpha_i \, dx_i^v \right) \tag{39}$$

$$d^\mu \alpha = \sum_{i=1}^{n} \sum_{j=1}^{n} dx_j^\mu \wedge \sum_{k=0}^{\infty} \binom{\mu}{k} \left( \frac{\partial^{\mu-k}}{(\partial x_j)^{\mu-k}} \alpha_j \right) \left( \frac{\partial^k}{\partial x_j^k} dx_i^v \right) \tag{40}$$

In the last sum of equation (40) 'k' takes on only whole number values hence

$$\frac{\partial^k}{\partial x_j^k}\left(dx_i^v\right) = 0 \quad \forall \; k \geq 1. \tag{41}$$

This reduces equation (40) to the following

$$d^\mu \alpha = \sum_{i=1}^{n} \sum_{j=1}^{n} dx_j^\mu \wedge dx_i^v \binom{\mu}{0} \left( \frac{\partial^\mu}{(\partial x_j)^\mu} \alpha_i \right). \tag{42}$$

Since $dx_j^\mu$ and $dx_i^v$ are linearly independent, provided $\mu \neq v$ or $i \neq j$, $d^\mu \alpha = 0$ if and only if

$$\frac{\partial^\mu}{(\partial x_j)^\mu} \alpha_i = 0. \tag{43}$$

In other words $\alpha_i \in Ker(\partial_j^\mu)$. For the special case of $\mu = \nu$ the symmetry from equation (29) can be used to obtain

$$\frac{\partial^\nu}{(\partial x_i)^\nu} \alpha_j + (-1)\frac{\partial^\nu}{(\partial x_j)^\nu} \alpha_i = 0. \tag{44}$$

For $\nu = 1$ the usual result from exterior calculus is recovered.

$$\frac{\partial}{\partial x_i} \alpha_j - \frac{\partial}{\partial x_j} \alpha_i = 0 \tag{45}$$

## VI. TRANSITION TO CURVILINEAR COORDINATES

When coordinate transformation rules are worked out for exterior or tensor calculus (see for example Refs. 5 and 6) the following construction can be used. Let $\{x_i\}$ and $\{y_i\}$ be two coordinate systems with a one to one mapping between them in some neighborhood of $P \in E^n$. Take $\{x_i\}$ to again be Cartesian coordinates and $\{y_i\}$ to be curvilinear coordinates. Assume the $\{x_i\}$ can be written smoothly in terms of the $\{y_i\}$.

$$x_i = x_i(y) \tag{46}$$

The exterior derivative is then applied to the above expression giving the following.

$$dx_i = dx_i(y) \tag{47}$$

$$\sum_{k=1}^{n} dx_k \frac{\partial x_i}{\partial x_k} = \sum_{l=1}^{n} dy_l \frac{\partial x_i}{\partial y_l} \tag{48}$$

$$\sum_{k=1}^{n} dx_k \delta_i^k = \sum_{l=1}^{n} dy_l \frac{\partial x_i}{\partial y_l} \tag{49}$$

$$dx_i = \sum_{l=1}^{n} dy_l \frac{\partial x_i}{\partial y_l} \tag{50}$$

This is very straightforward and can be adapted to the fractional form case. In the two coordinate systems the fractional exterior derivative $d^v$ takes the following forms

$$d^v = \sum_{i=1}^{n} dx_i^v \frac{\partial^v}{\left(\partial(x_i - a_i)\right)^v}, \tag{51}$$

and

$$d^v = \sum_{i=1}^{n} dy_i^v \frac{\partial^v}{\left(\partial(y_i - \tilde{a}_i)\right)^v}. \tag{52}$$

Where $a_i$ is the initial point of the derivative in the Cartesian system and $\tilde{a}_i$ is the same point but in the curvilinear coordinates. Recall that $x_i$ is in the kernel for the operator $\partial/\partial x_k$ for $k \neq i$, and when $k = i$ the result is one. For the fractional case the same type of object is needed. Consider a function $\alpha_k$ that maps points in $E^n$ into the complex numbers

$$\alpha_k = \frac{\Gamma(1)}{\Gamma(\nu+1)} \left( \prod_{\substack{i=1 \\ i \neq k}}^{n} (x_i - a_i) \right)^{\nu-m} (x_k - a_k)^\nu . \tag{53}$$

$a_i$ is the initial point for the fractional derivative. The function $\alpha_k$ was chosen so that it would be in the kernel of $\dfrac{\partial^\nu}{(\partial(x_i - a_i))^\nu}$ for $i \neq k$, and for $i = k$

$$\frac{\partial^\nu \alpha_k}{(\partial(x_k - a_k))^\nu} = 1 . \tag{54}$$

If the fractional exterior derivative is applied to $\alpha_k$ in the two different coordinate systems the following coordinate transformation rule can be obtained (where all quantities on the right hand side of (55) must be expressed in terms of the $\{y_i\}$ coordinates).

$$dx_k^\nu = \sum_{i=1}^{n} \frac{dy_i^\nu}{\Gamma(\nu+1)} \frac{\partial^\nu}{(\partial(y_i - \tilde{a}_i))^\nu} \left( \left( \prod_{\substack{i=1 \\ i \neq k}}^{n} (x_i - a_i) \right)^{\nu-m} (x_k - a_k)^\nu \right) \tag{55}$$

Note that for $\nu = m = 1$ the usual coordinate transformation rule is recovered. The coordinate transformation matrix for the fractional forms will be denoted by

$$J_i^k(x,y,\nu) = \frac{1}{\Gamma(\nu+1)} \frac{\partial^\nu}{(\partial(y_i - \tilde{a}_i))^\nu} \left( \left( \prod_{\substack{j=1 \\ j \neq k}}^{n} (x_j - a_j) \right)^{\nu-m} (x_k - a_k)^\nu \right), \tag{56}$$

$$dx_k^\nu = \sum_{j=1}^{n} dy_i^\nu J_i^k(x,y,\nu) . \tag{57}$$

A fractional form,

$$A = \sum_{k=1}^{n} A_k(x)\, dx_k^\nu \qquad (58)$$

written in the $\{x_i\}$ coordinates would be transformed into the $\{y_i\}$ coordinates according to

$$A = \sum_{k=1}^{n}\sum_{i=1}^{n} A_k(x(y)) J_i^k(x,y,\nu)\, dy_i^\nu . \qquad (59)$$

Reversing the coordinate transformation yields

$$A = \sum_{k=1}^{n}\sum_{i=1}^{n}\sum_{j=1}^{n} A_k(x) J_i^k(x,y,\nu) J_j^i(y,x,\nu)\, dx_j^\nu , \qquad (60)$$

$$\Rightarrow \delta_j^k = \sum_{i=1}^{n} J_j^i(y,x,\nu) J_i^k(x,y,\nu) . \qquad (61)$$

As an example consider the coordinate transformation for two dimensional Cartesian coordinates to polar coordinates. The initial point for the fractional derivatives is taken to be the origin.

$$x_1 = r\cos(\theta) \text{ and } x_2 = r\sin(\theta) \qquad (62)$$

The coordinate transformations for the fractional differentials are then

$$dx_1^v = \frac{\Gamma(2v-m+1)}{\Gamma(v+1)\Gamma(v-m+1)} \frac{\cos^v(\theta)}{\sin^{m-v}(\theta)} r^{v-m} dr^v + \frac{r^{2v-m}}{\Gamma(v+1)} \frac{\partial^v}{(\partial\theta)^v}\left(\frac{\cos^v(\theta)}{\sin^{m-v}(\theta)}\right) d\theta^v, \quad (63)$$

$$dx_2^v = \frac{\Gamma(2v-m+1)}{\Gamma(v+1)\Gamma(v-m+1)} \frac{\sin^v(\theta)}{\cos^{m-v}(\theta)} r^{v-m} dr^v + \frac{r^{2v-m}}{\Gamma(v+1)} \frac{\partial^v}{(\partial\theta)^v}\left(\frac{\sin^v(\theta)}{\cos^{m-v}(\theta)}\right) d\theta^v. \quad (64)$$

For $v = m = 1$ the transformation equations from exterior calculus are recovered.

$$dx_1 = \cos(\theta)dr - r\sin(\theta)d\theta \quad (65)$$

$$dx_2 = \sin(\theta)dr + r\cos(\theta)d\theta \quad (66)$$

Having found the coordinate transformation rule a metric for $F(v,1,n)$ can be constructed just as is done in exterior calculus (see page 46 of Lovelock and Rund [6])

$$g_{ij}(y,v) = \sum_{k=1}^{n} J_i^k(x,y,v) J_j^k(x,y,v). \quad (67)$$

Which can be used to give a fractional line element

$$ds^{2v} = \sum_{i,j=1}^{n} g_{ij}(y,v) dy_i^v \otimes dy_j^v \quad (68)$$

or,

$$ds^v = \sqrt{\sum_{i,j=1}^{n} g_{ij}(y,v) dy_i^v \otimes dy_j^v}. \quad (69)$$

Where $\otimes$ is the symmetric product for coordinate differentials.

## VII. CONCLUSION

In this paper a natural extension of the exterior derivative to a fractional exterior derivative was considered. It was found to generate some new vector spaces, both finite and infinite in dimension. Of particular interest is the observation that at each point $P \in E^n$ there are an infinite number of n dimensional vector spaces. Due to their similarity with tangent spaces, perhaps an appropriate name for these new vector spaces should be fractional tangent spaces. Notions of closed and exact were also defined for these fractional form spaces. Integrability and closure conditions were investigated for the special case of $F(v,1,n)$. The results produced were found to reduce to the standard results from exterior calculus when the order of the fractional exterior derivative was set equal to one. Coordinate transformation rules were also worked out for $F(v,1,n)$. The transformation rules are somewhat more complicated than for traditional exterior calculus. They do however reduce to the usual formula when the order of the coordinate differentials are set equal to one. Coordinate transformation rules give rise to a metric for $F(v,1,n)$. Properties of the metric, such as its associated covariant derivative, will be investigated in a later paper.

**ACKNOWLEDGMENT** One of the authors, M. Naber, would like to thank P. Dorcey for helpful comments and a critical reading of the paper. The authors also wish to thank the referee for several useful comments and corrections.